\documentclass[a4,11pt]{amsart}
\usepackage{amssymb}
\usepackage{amsmath}
\usepackage{amsfonts}
\usepackage{amsmath,amssymb,pstricks,pst-plot,pst-node,psfrag,graphicx,pst-grad,amsfonts}
\usepackage{float}

\usepackage{amssymb}
\usepackage{hyperref}
\usepackage{amsmath,amssymb,pstricks,pst-plot,pst-node,psfrag,graphicx,pst-grad,amsfonts}
\usepackage{enumerate}

\begin{document}

\title{The McDonald Modified Weibull Distribution: Properties and Applications}
%\author[F. Merovci$^{\ast }$ , I.Elbatal] \\

\author[Faton Merovci]{Faton Merovci  }
\address{Faton Merovci
\newline \indent Department of Mathematics,
\newline \indent University of Prishtina "Hasan Prishtina",
\newline \indent Republic of Kosovo}

\email{fmerovci@yahoo.com}

\author[Ibrahim Elbatal]{Ibrahim Elbatal  }
\address{Ibrahim Elbatal
\newline \indent Institute of Statistical Studies and Research,
\newline \indent Department of Mathematical Statistics,
\newline \indent Cairo University}

\email{i\_elbatal@staff.cu.edu.eg}

\maketitle

\begin{abstract}
A six parameter distribution so-called the McDonald modified Weibull
distribution is defined and studied. The new distribution contains, as
special submodels, several important distributions discussed in the
literature, such as the beta modified Weibull, Kumaraswamy modified
Weibull, McDonald Weibull and modified Weibull distribution,among others.
The new distribution can be used effectively in the analysis of survival
data since it accommodates monotone, unimodal and bathtub-shaped hazard
functions. We derive the moments.We propose the method of maximum
likelihood for estimating the model parameters and obtain the observed information matrix. A real data set
is used to illustrate the importance and flexibility of the new distribution.

Keywords: McDonald  modified weibull distribution , Moments, Maximum likelihood .
\end{abstract}

\section{Introduction}

The modified Weibull (MW) distribution is one of the most important
distributions in lifetime modeling, and some well-known distributions are
special cases of it. This distribution was introduced by Lai, Xie, and
Murthy (2003) to which we refer the reader for a detailed discussion as well
as applications of the MW distribution (in particular, the use of a real
data set representing failure times to illustrate the modeling and
estimation procedure). Also Sarhan and Zaindin (2008) introduced the
modified Weibull distribution . It can be used to describe several
reliability models. It has three parameters, two scale and one shape
parameters. Recently, Carrasco, Ortega, and Cordeiro (2008), Ortega,
Cordeiro, and Carrasco (2011) extended the MW distribution by adding another
shape parameter and introducing a four parameter generalized MW (GMW) and
log-GMW (LGMW). In Section 8 of Carrasco et al. (2008) two applications of
GMW in serum-reversal data and radiotherapy data are presented to which we
refer the interested reader for details. In Section 6 of Ortega et al.
(2011), an application of LGMW in survival times for the golden shiner data
is presented. Although in many applications an increase in the number of
parameters provides a more suitable model, in the characterization problem a
lower number of parameters (without affecting the suitability of the model)
is mathematically more appealing (see Glanzel and Hamedani, 2001),
especially in the MW case which already has a shape parameter. So, we
restrict our attention to the MW and log-MW (LMW) distributions. In the
applications where the underlying distribution is assumed to be MW (or LMW),
the investigator needs to verify that the underlying distribution is in fact
the MW (or LMW). To this end the investigator has to rely on the
characterizations of these distributions and determine if the corresponding
conditions are satisfied. Thus, the problem of characterizing the MW (or
LMW) become essential.

A random variable $X$ is said to have modified Weibull distribution if
cumulative distribution function(cdf) is%
\begin{equation}\label{eq1.1}
F(x)=1-e^{-\alpha x-\gamma x^{\beta }},x\geq 0,
\end{equation}%
where $\beta >0$, $\alpha ,\gamma \geq 0$, such that $\alpha +\gamma >0
$. Here $\alpha $ is a scale parameter, while $\gamma $ and $\beta $ are
shape parameters. The corresponding probability density function(pdf) is%
\begin{equation}\label{eq1.2}
f(x)=\left( \alpha +\gamma \beta x^{\beta -1}\right) e^{-\alpha x-\gamma
x^{\beta }},
\end{equation}%
and the hazard function is given by%
\begin{align}\label{eq1.3}
h(x) &=\frac{f(x)}{1-F(x)}=\frac{\left( \alpha +\gamma \beta x^{\beta
-1}\right) e^{-\alpha x-\gamma x^{\beta }}}{e^{-\alpha x-\gamma x^{\beta }}}\notag \\
&=\left( \alpha +\gamma \beta x^{\beta -1}\right) .
\end{align}%
One can easily verify from \eqref{eq1.3} that: 
\begin{enumerate}[i)] 
\item The hazard function is
constant when $\beta =1$; 
\item  when $\beta <1$, the hazard
function is decreasing; and 
\item the hazard function will be increasing if $%
\beta >1$.
\end{enumerate}

Consider an arbitrary parent cdf $G(x)$. The probability density function
(pdf) $f(x)$ of the new class of distributions called the Mc-Donald
generalized distributions (denoted with the prefix " Mc" for short) is
defined by%
\begin{equation}\label{eq1.4}
f(x,a,b,c)=\frac{c}{B(a,b)}g(x)G^{ac-1}(x)\left[ 1-G^{c}(x)\right] ^{b-1},
\end{equation}%
where $a>0,b>0$ and $c>0$ are additional shape parameters . ( See Corderio
et al. (2012) for additional details). Note that $g(x)$ is the pdf of parent
distribution ,$g(x)=\frac{dG(x)}{dx}$. Introduction of this additional shape
parameters is specially to introduce skewness. Also, this allows us to vary
tail weight. It is important to note that for $c=1$ we obtain a sub-model of
this generalization which is a beta generalization ( see Eugene et al.(
2002)) and for $a=1$, we have the Kumaraswamy (Kw), [Kumaraswamy generalized
distributions ( see Cordeiro and Castro, (2010)). For random variable $X$
with density function \eqref{eq1.4}, we write $X\sim Mc-G(a,b,c)$. The probability
density function \eqref{eq1.4} will be most tractable when $G(x)$ and $g(x)$ have
simple analytic expressions. The corresponding cumulative function for this
generalization is given by%
\begin{equation}\label{eq1.5}
F(x,a,b,c)=I_{G^{c}(x)}(a,b)=\frac{1}{B(a,b)}\int\limits_{0}^{G(x)^{c}}w^{(1-a)}(1-w)^{b-1}dw,
\end{equation}%
where $I_{y}(a,b)=\frac{1}{B(a,b)}\int\limits_{0}^{y}w^{(1-a)}(1-w)^{b-1}dw$ denotes the incomplete beta
function ratio (Gradshteyn \& Ryzhik, 2000). Equation \eqref{eq1.5} can also be
rewritten as follows%
\begin{equation}\label{eq1.6}
F(x,a,b,c)=\frac{G(x)^{ac}}{aB(a,b)}\text{ }_{2}F_{1}(a,1-b;a+1;G(x)^{c}),
\end{equation}
where
\begin{equation*}
_{2}F_{1}(a,b;c;x)=B(b,c-b)^{-1}\int\limits_{0}^{1}\frac{t^{b-1}\left(
1-t\right) ^{c-b-1}}{\left( 1-tx\right) ^{a}}dt
\end{equation*}%
is the well-known hypergeometric functions which are well established in the
literature ( see,Gradshteyn and Ryzhik (2000)).

\medskip

Some mathematical properties of the cdf $F(x)$ for any Mc-G distribution
defined from a parent $G(x)$ in equation \eqref{eq1.5}, could, in principle, follow
from the properties of the hypergeometric function, which are well
established in the literature (Gradshteyn and Ryzhik, 2000, Sec. 9.1). One
important benefit of this class is its ability to skewed data that cannot
properly be fitted by many other existing distributions. Mc-G family of
densities allows for higher levels of flexibility of its tails and has a lot
of applications in various fields including economics, finance, reliability,
engineering, biology and medicine.\medskip

Figure 1 and figure 2 illustrates some of the possible shapes of the pdf and cdf of McMW distribution  for selected values of the parameters $\alpha, \beta. \gamma, a, b $ and $c$, respectively.

\begin{figure}[H]
\centering
 \vspace*{-20mm}
  \includegraphics[width=8.0cm]{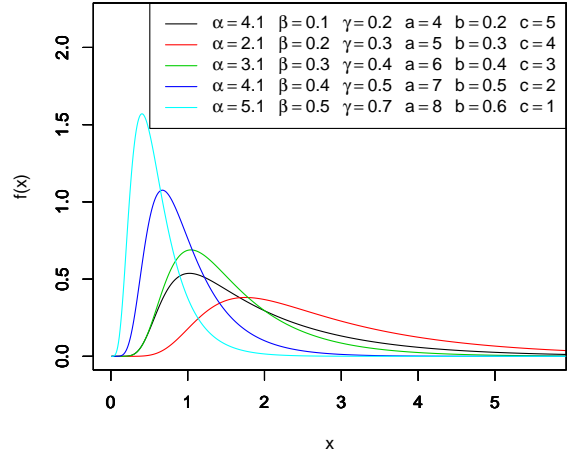}
 \caption{The pdf's of various McMW distributions.\label{fig1.pdf}}
\end{figure}
\begin{figure}[H]
\centering
 \vspace*{-20mm}
 \includegraphics[width=8.0cm]{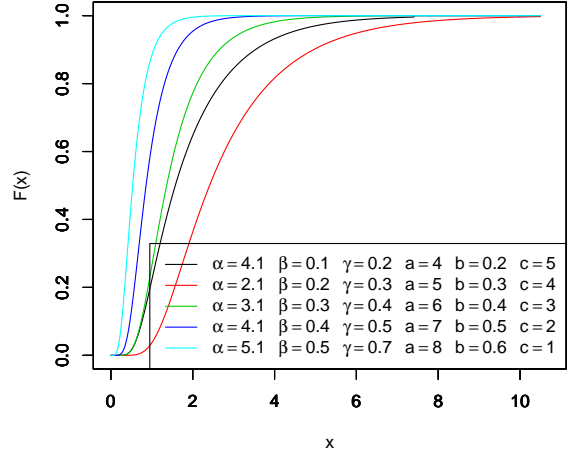}
 \caption{The cdf's of various McMW distributions.\label{fig2.pdf}}
 \end{figure}

The hazard function (hf) and reverse hazard functions (rhf) of the Mc-G
distribution are given by%
\begin{equation}\label{eq1.7}
h(x)=\frac{f(x)}{1-F(x)}=\frac{cg(x)G^{ac-1}(x)\left[ 1-G^{c}(x)\right]
^{b-1}}{B(a,b)\left\{ 1-I_{G^{c}(x)}(a,b)\right\} },
\end{equation}%
and%
\begin{equation*}
\tau (x)=\frac{f(x)}{F(x)}=\frac{cg(x)G^{ac-1}(x)\left[ 1-G^{c}(x)\right]
^{b-1}}{B(a,b)\left\{ I_{G^{c}(x)}(a,b)\right\} }
\end{equation*}%
respectively.

Figure 3 and 4 illustrates some of the possible shapes of the hazard rate function  and survival function of McMW distribution  for selected values of the parameters $\alpha, \beta. \gamma, a, b $ and $c$.

\begin{figure}[H]
\centering
 \vspace*{-20mm}
  \includegraphics[width=8.0cm]{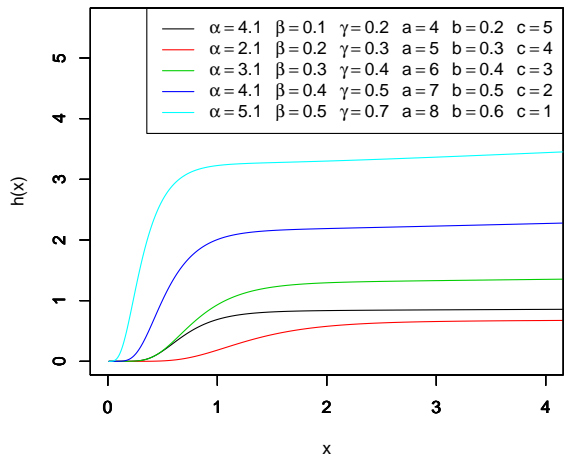}
 \caption{The hazard function of various McMW distributions.\label{fig3.pdf}}
 \end{figure}

\begin{figure}[H]
\centering
 \vspace*{-20mm}
  \includegraphics[width=8.0cm]{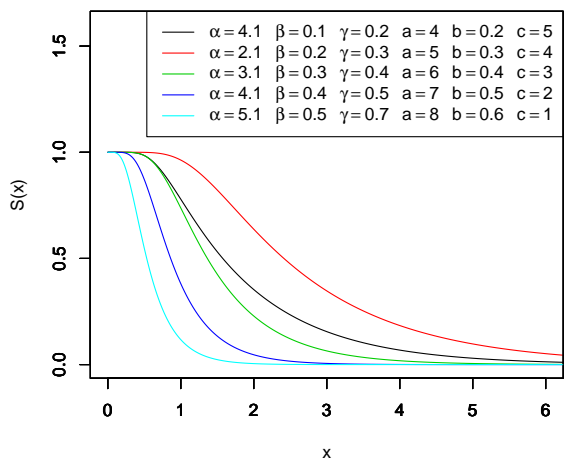}
 \caption{The survival function of various McMW distributions.\label{fig4.pdf}}
 \end{figure}

Recently Cordeiro et al. (2012) introduced the The McDonald
Normal Distribution. Also, Francisco et al. (2012) proposed a new
distribution, called the McDonald gamma distribution.

The rest of the paper is organized as follows. In Section 2, we demonstrate
that the $McMW$ density function can be expressed as a linear combination of
the modified Weibull distribution. This result is important to provide
mathematical properties of the $McMW$ model directly from those properties
of the modified Weibull distribution in Section 3. In Section 4 we discuss
some important statistical properties of the $McMW$ distribution including
quantile function, moments , moment generating function. The distribution
of the order statistics is expressed in Section 5 . Maximum
likelihood estimates of the parameters index to the distribution are
discussed in Section 6. Section 7 provides applications
to  real data sets. Section 8 ends with some conclusions.

\section{McDonald Modified Weibull Distribution}

In this section we studied the McDonald modified Weibull $(McMW)$
distribution and the sub-models of this distribution. Using $G(x)$ and $g(x)$
in \eqref{eq1.4}) to be the cdf and pdf of \eqref{eq1.1} and \eqref{eq1.2}. The pdf of the $McMW$
distribution is given by%
\begin{align}\label{eq2.1}
f(x,\varphi )&=\frac{c}{B(a,b)}\left( \alpha +\gamma \beta x^{\beta
-1}\right) e^{-\alpha x-\gamma x^{\beta }}\left[ 1-e^{-\alpha x-\gamma
x^{\beta }}\right] ^{ac-1}\notag\\
&\cdot\left[ 1-\left( 1-e^{-\alpha x-\gamma x^{\beta
}}\right) ^{c}\right] ^{b-1},x>0.
\end{align}%
where $\varphi =(\alpha ,\gamma ,\beta ,a,b,c)$. The corresponding cdf of
the $McMW$ distribution is given by%
\begin{align}\label{eq2.2}
F(x) &=I_{G^{c}(x)}(a,b)=\frac{1}{B(a,b)}\int
\limits_{0}^{G(x)^{c}}w^{(1-a)}(1-w)^{b-1}dw  \notag \\
&=\frac{1}{B(a,b)}\int\limits_{0}^{\left( 1-e^{-\alpha x-\gamma x^{\beta
}}\right) ^{c}}w^{(1-a)}(1-w)^{b-1}dw  \notag \\
&=I_{\left( 1-e^{-\alpha x-\gamma x^{\beta }}\right) ^{c}}(a,b),
\end{align}%
also, the cdf can be written as follows
\begin{equation}\label{eq2.3}
F(x)=\frac{\left( 1-e^{-\alpha x-\gamma x^{\beta }}\right) ^{ac}}{aB(a,b)}%
\text{ }_{2}F_{1}(a,1-b;a+1;\left[ 1-e^{-\alpha x-\gamma x^{\beta }}\right]
^{c},
\end{equation}%
where $_{2}F_{1}(a,b;c;x)=B(b,c-b)^{-1}\int\limits_{0}^{1}\frac{%
t^{b-1}\left( 1-t\right) ^{c-b-1}}{\left( 1-tx\right) ^{a}}dt.$

The hazard rate function and reversed hazard rate function of the new
distribution are given by%
\begin{align}\label{eq2.4}
h(x) &=\frac{f(x)}{1-F(x)}  \notag \\
&=\frac{c\left( \alpha +\gamma \beta x^{\beta -1}\right) e^{-\alpha
x-\gamma x^{\beta }}\left[ 1-e^{-\alpha x-\gamma x^{\beta }}\right] ^{ac-1}%
\left[ 1-\left( 1-e^{-\alpha x-\gamma x^{\beta }}\right) ^{c}\right] ^{b-1}}{%
B(a,b)\left\{ 1-I_{\left( 1-e^{-\alpha x-\gamma x^{\beta }}\right)
^{c}}(a,b)\right\} },
\end{align}
and
\begin{align}\label{eq2.5}
\tau (x) &=\frac{f(x)}{F(x)}  \notag \\
&=\frac{c\left( \alpha +\gamma \beta x^{\beta -1}\right) e^{-\alpha
x-\gamma x^{\beta }}\left[ 1-e^{-\alpha x-\gamma x^{\beta }}\right] ^{ac-1}%
\left[ 1-\left( 1-e^{-\alpha x-\gamma x^{\beta }}\right) ^{c}\right] ^{b-1}}{%
B(a,b)\left\{ I_{\left( 1-e^{-\alpha x-\gamma x^{\beta }}\right)
^{c}}(a,b)\right\} },
\end{align}
respectively.

\subsection{Submodels}

The McDonald modified Weibull distribution is very flexible model that
approaches to different distributions when its parameters are changed. The $McMW$ distribution contains as special- models the following well known
distributions. If $X$ is a random variable with pdf \eqref{eq2.1} or cdf \eqref{eq2.2} we
use the notation $X\sim McMW$ $(\alpha ,\gamma ,\beta ,a,b,c)$ then
we have the following cases.
\begin{enumerate}
\item  For $c=1$ , then \eqref{eq2.1} reduces to the beta modified Weibull $(BMW)$
distribution.
\item For $a=1$ we get the kumaraswamy modified weibull $(KMW)$ distribution.

\item  For $\alpha =1$ , then \eqref{eq2.1} becomes McDonald Weibull $(McW)$
distribution.
\item  For $\beta =2$ , we get the McDonald Linear Failure Rate $(McLFR)$
distribution.
\item   For $\alpha =0$ and $\beta =2$ , then \eqref{eq2.1} becomes McDonald Rayleigh $(McR)$ distribution.
\item The McDonald Exponential $(McE)$ distribution arises as a special case of
$McMW$  by taking $\gamma =0.$
\item For $\alpha =0$ , $\beta =2$  and $c=1$, then \eqref{eq2.1} becomes beta
Rayleigh $(BR)$ distribution.
\item For $\beta =2$ and $c=1$ , we get the beta Linear Failure Rate $(BLFR)$
distribution.
\item Applying $a=b=c=1$ we can obtain the modified Weibull distribution.
\end{enumerate}

 The flexibility of the McDonald modified Weibull distribution is explained
in Table (1). The subject distribution includes as special cases the
McDonald modified Weibull($McMW$), beta modified Weibull  (BMW),
Kumaraswamy modified Weibull (KMW), McDonald, McDonald exponential (McE),
McDonald linear failure rate (McLFR), modified Weibull (MW), Rayleigh (R),
Exponential and Linear failure rate distributions.

\begin{table}[hbt]
 \caption{\label{Tab1} Sub-models of McMW Distribution.}
 \begin{small}
 \begin{center}
 \begin{tabular}{|lllllll|}
 \hline
$Distribution$ $(McMW)$ & $\alpha $ & $\gamma $ & $\beta $ & $a$ & $b$ & $c$
\\ \hline
$BMW$ & $\alpha $ & $\gamma $ & $\beta $ & $a$ & $b$ & $1$ \\ \hline
$KMW$ & $\alpha $ & $\gamma $ & $\beta $ & $1$ & $b$ & $1$ \\ \hline
$McW$ & $0$ & $\gamma $ & $\beta $ & $a$ & $b$ & $c$ \\ \hline
$McLFR$ & $\alpha $ & $\gamma $ & $2$ & $a$ & $b$ & $c$ \\ \hline
$McR$ & $0$ & $\gamma $ & $2$ & $a$ & $b$ & $c$ \\ \hline
$McE$ & $\alpha $ & $0$ & $0$ & $a$ & $b$ & $c$ \\ \hline
$BR$ & $0$ & $\gamma $ & $2$ & $a$ & $b$ & $1$ \\ \hline
$BLFR$ & $\alpha $ & $\gamma $ & $2$ & $a$ & $b$ & $1$ \\ \hline
$MW$ & $\alpha $ & $\gamma $ & $\beta $ & $1$ & $1$ & $1$ \\ \hline
 \end{tabular}
 \end{center}
 \end{small}
\end{table}

\section{Expansion of Distribution}

In this section,we present a series expansion of the $McMW$ \ cdf and pdf.
distribution depending if the parameter $b>0$ is real non- integer or
integer. First, if $\left\vert z\right\vert <1$ and $b>0$ is real non-
integer, we have%
\begin{equation}\label{eq3.1}
(1-z)^{b-1}=\sum\limits_{j=0}^{\infty }(-1)^{j}\binom{b-1}{j}%
z^{j}=\sum\limits_{j=0}^{\infty }\frac{(-1)^{j}\Gamma (b)}{j!\Gamma (b-j)}%
z^{j}.
\end{equation}%
Using the expansion \eqref{eq3.1} in \eqref{eq2.2}, the cdf of the $McMW$ distribution
becomes%
\begin{align*}
F(x) &=\frac{1}{B(a,b)}\int\limits_{0}^{\left( 1-e^{-\alpha x-\gamma
x^{\beta }}\right) ^{c}}w^{(1-a)}(1-w)^{b-1}dw \\
&=\frac{\Gamma (b)}{B(a,b)}\sum\limits_{j=0}^{\infty }\frac{(-1)^{j}}{%
j!\Gamma (b-j)}\int\limits_{0}^{G(x)^{c}}w^{a+j-1}dw \\
&=\sum\limits_{j=0}^{\infty }\frac{(-1)^{j}\Gamma (b)}{B(a,b)j!\Gamma
(b-j)(a+j)}\left[ G(x,\alpha ,\gamma ,\beta )\right] ^{c(a+j)} \\
&=\sum\limits_{j=0}^{\infty }q_{j}\text{ }G(x,\alpha c(a+j),\gamma ,\beta )
\end{align*}%
If $b>0$ is an integer, then%
\begin{equation}\label{eq3.2}
F(x)=\sum\limits_{j=0}^{b-1}q_{j}G(x,\alpha c(a+j),\gamma ,\beta ).
\end{equation}%
Similarly, if $b>0$ is real non- integer the pdf is given by
\begin{equation}\label{eq3.3}
f(x)=\sum\limits_{j=0}^{\infty }q_{j}g(x,\alpha c(a+j),\gamma
,\beta ),
\end{equation}%
and%
\begin{equation}\label{eq3.4}
f(x)=\sum\limits_{j=0}^{b-1}q_{j}g(x,\alpha c(a+j),\gamma ,\beta ),
\end{equation}%
for $b>0$ is an integer. Where $q_{j}=\frac{(-1)^{j}\Gamma (b)}{%
B(a,b)j!\Gamma (b-j)(a+j)}$ are constants such that $\sum\limits_{j=0}^{%
\infty }q_{j}=1$ and $G(x,\alpha c(a+j),\gamma ,\beta )$ is a finite mixture
of modified Weibull distribution with scale parameter $\alpha c(a+j)$ and $%
\gamma ,\beta $ are shape parameters.

\section{Statistical Properties}

In this section we discuss the statistical properties of the McDonald
modified Weibull distribution, in particular, quantile function, moment and
moment generating function.

\subsection{Quantile function}

The $McQL$ quantile function , say $Q(u)=F^{-1}(u)$, is straightforward
to be computed by inverting \eqref{eq1.8}, we have%
\begin{equation}\label{eq4.1}
\gamma x_{q}^{\beta }+\alpha x_{q}^{\theta }+\ln \left[ 1-Q_{(a,b)}(u)^{%
\frac{1}{c}})\right] =0,
\end{equation}%
we can easily generate $X$ by taking $u$ as a uniform random variable in $%
(0,1)$.

\subsection{Moments}

In this subsection we discuss the $r_{th}$ moment for $McMW$ distribution.
Moments are necessary and important in any statistical analysis, especially
in applications. It can be used to study the most important features and
characteristics of a distribution (e.g., tendency, dispersion, skewness and
kurtosis).\medskip \newline

\textbf{Theorem (4.1)}. If $X$ has $McMW$ $(\varphi ,x)$ $,\varphi =(\alpha ,\gamma ,\beta ,a,b,c)$
then the $k_{th}$ moment of $X$ is given by the following%
\begin{equation}\label{eq4.2}
\mu _{r}(x)=w_{j,m,s}\left[ \frac{\alpha \Gamma (k+\beta s+1)}{\left[
m(\alpha +1)\right] ^{k+\beta s+1}}+\frac{\gamma \beta \Gamma (k+\beta (s+1))%
}{\left[ m(\alpha +1)\right] ^{k+\beta (s+1)}}\right] .
\end{equation}%

\textbf{Proof}:

Let $X$ be a random variable with density function \eqref{eq2.1}. The $k_{th}$
ordinary moment of the $McMW$ distribution is given by%
\begin{align}\label{eq4.3}
\mu _{k}^{^{\prime }}(x) &=E(X^{k)}=\int\limits_{0}^{\infty
}x^{k}f(x,\varphi )dx  \notag \\
&&  \notag \\
&=\frac{c}{B(a,b)}\int\limits_{0}^{\infty }x^{k}\left( \alpha +\gamma
\beta x^{\beta -1}\right) e^{-\alpha x-\gamma x^{\beta }}\left[ 1-e^{-\alpha
x-\gamma x^{\beta }}\right] ^{ac-1}\left[ 1-\left( 1-e^{-\alpha x-\gamma
x^{\beta }}\right) ^{c}\right] ^{b-1}dx.
\end{align}%
Setting
\begin{equation}\label{eq4.4}
\left[ 1-\left( 1-e^{-\alpha x-\gamma x^{\beta }}\right) ^{c}\right]
^{b-1}=\sum\limits_{j=0}^{\infty }(-1)^{j}\binom{b-1}{j}\left(
1-e^{-\alpha x-\gamma x^{\beta }}\right) ^{cj},
\end{equation}%
so%
\begin{equation*}
\mu _{k}^{^{\prime }}(x)=\frac{c}{B(a,b)}\sum\limits_{j=0}^{\infty }(-1)^{j}%
\binom{b-1}{j}\int\limits_{0}^{\infty }x^{k}\left( \alpha +\gamma \beta
x^{\beta -1}\right) e^{-\alpha x-\gamma x^{\beta }}\left[ 1-e^{-\alpha
x-\gamma x^{\beta }}\right] ^{c(a+j)-1}
\end{equation*}
but
\begin{equation}\label{eq4.5}
\left[ 1-e^{-\alpha x-\gamma x^{\beta }}\right] ^{c(a+j)-1}=\sum\limits_{k=0}^{\infty }(-1)^{m}\binom{c(j+a)-1}{m}e^{-m\alpha x-m\gamma
x^{\beta }},
\end{equation}%
then%
\begin{align}\label{eq4.6}
\mu _{k}^{^{\prime }}(x) &=\frac{c}{B(a,b)}\sum\limits_{j=0}^{\infty
}\sum\limits_{m=0}^{\infty }(-1)^{j+m}\binom{b-1}{j}\binom{c(j+a)-1}{m}
\notag \\
&\times \int\limits_{0}^{\infty }x^{k}\left( \alpha +\gamma \beta
x^{\beta -1}\right) e^{-m(\alpha +1)x-m(\gamma +1)x^{\beta }}dx
\end{align}%
using the following expansion of $e^{-m(\gamma +1)x^{\beta }}$ given by%
\begin{equation*}
e^{-m(\gamma +1)x^{\beta }}=\sum\limits_{s=0}^{\infty }\frac{(-1)^{s}}{s!}%
\left[ m(\gamma +1)\right] ^{s}x^{\beta s}
\end{equation*}%
thus equation \eqref{eq4.6} takes the following form
\begin{align}\label{eq4.7}
\mu _{k}^{^{\prime }}(x) &=\frac{c\left[ m(\gamma +1)\right] ^{s}}{B(a,b)s!}%
\sum\limits_{j,m,s=0}^{\infty }(-1)^{j+m+s}\binom{b-1}{j}\binom{c(j+a)-1}{%
m}  \notag \\
&\times \left\{ \alpha \int\limits_{0}^{\infty }x^{k+\beta
s}e^{-m(\alpha +1)x}dx+\gamma \beta \int\limits_{0}^{\infty }x^{k+\beta
(s+1)-1}e^{-m(\alpha +1)x}dx\right\}  \notag \\
&=w_{j,m,s}\left[ \frac{\alpha \Gamma (k+\beta s+1)}{\left[ m(\alpha +1)%
\right] ^{k+\beta s+1}}+\frac{\gamma \beta \Gamma (k+\beta (s+1))}{\left[
m(\alpha +1)\right] ^{k+\beta (s+1)}}\right]
\end{align}%
where%
\begin{equation*}
w_{j,m,s}=\frac{c\left[ m(\gamma +1)\right] ^{s}}{B(a,b)s!}%
\sum\limits_{j,m,s=0}^{\infty }(-1)^{j+m+s}\binom{b-1}{j}\binom{c(j+a)-1}{%
m}.
\end{equation*}%
Which completes the proof .\medskip \newline
Based on the first four moments of the $(McMW)$ distribution, the measures
of skewness $A(\varphi )$ and kurtosis $k(\varphi )$ of the $(McMW)$ distribution
can obtained as%
\begin{equation}\label{4.8}
A(\varphi )=\frac{\mu _{3}(\theta )-3\mu _{1}(\theta )\mu _{2}(\theta )+2\mu
_{1}^{3}(\theta )}{\left[ \mu _{2}(\theta )-\mu _{1}^{2}(\theta )\right] ^{%
\frac{3}{2}}},
\end{equation}%
and
\begin{equation}\label{eq4.9}
k(\varphi )=\frac{\mu _{4}(\theta )-4\mu _{1}(\theta )\mu _{3}(\theta )+6\mu
_{1}^{2}(\theta )\mu _{2}(\theta )-3\mu _{1}^{4}(\theta )}{\left[ \mu
_{2}(\theta )-\mu _{1}^{2}(\theta )\right] ^{2}}.
\end{equation}%
\textbf{Theorem (4.2):}\medskip \newline

If $X$ has the $McMW$ $(\alpha ,\gamma ,\beta ,a,b,c,x)$ then the the moment
generating function (mgf) of $X$ \ is given as follows%
\begin{equation}\label{eq4.10}
M_{X}(t)=w_{j,m,s}\left[ \frac{\alpha \Gamma (\beta s+1)}{\left[ m(\alpha
+1)-t\right] ^{\beta s+1}}+\frac{\gamma \beta \Gamma (\beta (s+1))}{\left[
m(\alpha +1)-t\right] ^{\beta (s+1)}}\right] .
\end{equation}%
\textbf{Proof:}

Starting with%
\begin{align}\label{eq4.11}
M_{X}(t) &=\int\limits_{0}^{\infty }e^{tx}f_{McMW}(\alpha ,\gamma ,\beta
,a,b,c,x)dx  \notag \\
&=w_{j,m,s}\int\limits_{0}^{\infty }(\alpha x^{\beta s}+\gamma \beta
x^{\beta (s+1)-1})e^{-x(m(\alpha +1)-t)}dx  \notag \\
&=w_{j,m,s}\left\{ \alpha \int\limits_{0}^{\infty }x^{\beta
s}e^{-x(m(\alpha +1)-t)}dx+\gamma \beta \int\limits_{0}^{\infty }x^{\beta
(s+1)-1}e^{-x(m(\alpha +1)-t)}dx\right\}  \notag \\
&=w_{j,m,s}\left[ \frac{\alpha \Gamma (\beta s+1)}{\left[ m(\alpha +1)-t%
\right] ^{\beta s+1}}+\frac{\gamma \beta \Gamma (\beta (s+1))}{\left[
m(\alpha +1)-t\right] ^{\beta (s+1)}}\right]
\end{align}%
Which completes the proof.

\section{Distribution of the order statistics}

In this section, we derive closed form expressions for the pdfs of the $%
r_{th}$ order statistic of the $(McMW)$ distribution, also, the measures of
skewness and kurtosis of the distribution of the $r_{th}$ order statistic in
a sample of size $n$ for different choices of $n;r$ are presented in this
section. Let $X_{1},X_{2},...,X_{n}$ be a simple random sample from $(McMW)$
distribution with pdf and cdf given by \eqref{eq2.1} and \eqref{eq2.3}, respectively.

Let $X_{1},X_{2},...,X_{n}$ denote the order statistics obtained from this
sample. We now give the probability density function of $X_{r:n}$, say $%
f_{r:n}(x,\varphi )$ and the moments of $X_{r:n}$ $,r=1,2,...,n$. Therefore,
the measures of skewness and kurtosis of the distribution of the $X_{r:n}$
are presented. The probability density function of $X_{r:n}$ is given by%
\begin{equation}\label{eq5.1}
f_{r:n}(x,\varphi )=\frac{1}{B(r,n-r+1)}\left[ F(x,\Phi )\right] ^{r-1}\left[
1-F(x,\Phi )\right] ^{n-r}f(x,\varphi )
\end{equation}%
where $F(x,\varphi )$ and $f(x,\varphi )$ are the cdf and pdf of the $(McMW)$
distribution given by \eqref{eq2.2}, \eqref{eq2.3}, respectively, and $B$ $(.,.)$ is the
beta function, since $0<F(x,\varphi )<1$, for $x>0$, by using the binomial
series expansion of $\left[ 1-F(x,\varphi )\right] ^{n-r}$, given by%
\begin{equation}\label{eq5.2}
\left[ 1-F(x,\varphi )\right] ^{n-r}=\sum\limits_{j=0}^{n-r}(-1)^{j}\binom{%
n-r}{j}\left[ F(x,\varphi )\right] ^{^{j}},
\end{equation}%
we have%
\begin{equation}\label{eq5.3}
f_{r:n}(x,\varphi )=\sum\limits_{j=0}^{n-r}(-1)^{j}\binom{n-r}{j}\left[
F(x,\Phi )\right] ^{r+j-1}f(x,\varphi ),
\end{equation}%
substituting from \eqref{eq2.2} and \eqref{eq2.3} into \eqref{eq5.3}), we can express the $k_{th}$
ordinary moment of the $r_{th}$ order statistics $X_{r:n}$  say $%
E(X_{r:n}^{k})$ as a liner combination of the $k_{th}$ moments of the $%
(McMW) $ distribution with different shape parameters. Therefore, the
measures of skewness and kurtosis of the distribution of $X_{r:n}$ can be
calculated.

\section{Maximum Likelihood Estimators}

In this section we consider the maximum likelihood estimators (MLE's) of $%
McMW$ $(\alpha ,\gamma ,\beta ,a,b,c,x)$ . Let $x_{1},...,$ $x_{n}$ be a
random sample of size $n$ from $McMW$ $(\alpha ,\gamma ,\beta ,a,b,c,x)$
then the likelihood function can be written as$e^{-\alpha x-\gamma x^{\beta
}}$%
\begin{align}\label{eq6.1}
L(\alpha ,\beta ,\theta ,\gamma ,\lambda ,x_{_{(i)}})
&=\prod\limits_{i=1}^{n}\left( \frac{c}{B(a,b)}\right)
\prod\limits_{i=1}^{n}\left( \alpha +\gamma \beta x_{i}^{\beta -1}\right)
\prod\limits_{i=1}^{n}e^{-\alpha \sum\limits_{i=1}^{n}x_{i}-\gamma
\sum\limits_{i=1}^{n}x_{i}^{\beta }}  \notag \\
&cdot \prod\limits_{i=1}^{n}\left[
1-e^{-\alpha x_{i}-\gamma x_{i}^{\beta }}\right] ^{ac-1} \prod\limits_{i=0}^{n}\left[ 1-\left( 1-e^{-\alpha x_{i}-\gamma
x_{i}^{\beta }}\right) ^{c}\right] ^{b-1}
\end{align}%
By accumulation taking logarithm of equation \eqref{eq6.1} , and the log- likelihood
function can be written as%
\begin{align}\label{eq6.2}
\ell=\log L &=n\log c+n\log \left[ \Gamma (a+b)\right] -n\log \left[ \Gamma (a)%
\right] -n\log \left[ \Gamma (b)\right]  \notag \\
&+\sum\limits_{i=1}^{n}\log \left( \alpha +\gamma \beta x_{i}^{\beta
-1}\right) -\alpha \sum\limits_{i=1}^{n}x_{i}-\gamma
\sum\limits_{i=1}^{n}x_{i}^{\beta }  \notag \\
&+(ac-1)\sum\limits_{i=1}^{n}\log \left[ 1-e^{-\alpha x_{i}-\gamma
x_{i}^{\beta }}\right]  \notag \\
&+(b-1)\sum\limits_{i=1}^{n}\log \left[ 1-\left( 1-e^{-\alpha
x_{i}-\gamma x_{i}^{\beta }}\right) ^{c}\right] .
\end{align}%
Computing the first partial derivatives of $\ell$ and setting the results
equal zeros, we get the likelihood equations as in the following form%
\begin{equation}\label{eq6.3}
\frac{\partial \ell}{\partial a}=n\psi (a+b)-n\psi
(a)+c\sum\limits_{i=1}^{n}\log \left[ 1-e^{-\alpha x_{i}-\gamma
x_{i}^{\beta }}\right] ,
\end{equation}%
\begin{equation}\label{eq6.4}
\frac{\partial \ell}{\partial b}=n\psi (a+b)-n\psi
(b)+\sum\limits_{i=1}^{n}\log \left[ 1-\left( 1-e^{-\alpha x_{i}-\gamma
x_{i}^{\beta }}\right) ^{c}\right] ,
\end{equation}%
\begin{align}\label{eq6.5}
\frac{\partial \ell}{\partial c} &=\frac{n}{c}+a\sum\limits_{i=1}^{n}%
\log \left[ 1-e^{-\alpha x_{i}-\gamma x_{i}^{\beta }}\right]  \notag \\
&-(b-1)\sum\limits_{i=1}^{n}\frac{\left( 1-e^{-\alpha x_{i}-\gamma
x_{i}^{\beta }}\right) ^{c}\log \left( 1-e^{-\alpha x_{i}-\gamma
x_{i}^{\beta }}\right) }{\left[ 1-\left( 1-e^{-\alpha x_{i}-\gamma
x_{i}^{\beta }}\right) ^{c}\right] },
\end{align}%
\begin{align}\label{eq6.6}
\frac{\partial \ell}{\partial \alpha } &=\sum\limits_{i=1}^{n}\frac{1}{%
\left( \alpha +\gamma \beta x_{i}^{\beta -1}\right) }-\sum%
\limits_{i=1}^{n}x_{i}+(ac-1)\sum\limits_{i=1}^{n}\frac{x_{i}e^{-\alpha
x_{i}-\gamma x_{i}^{\beta }}}{\left[ 1-e^{-\alpha x_{i}-\gamma x_{i}^{\beta
}}\right] }  \notag \\
&+c(b-1)\sum\limits_{i=1}^{n}\frac{x_{i}e^{-\alpha x_{i}-\gamma
x_{i}^{\beta }}\left( 1-e^{-\alpha x_{i}-\gamma x_{i}^{\beta }}\right) ^{c-1}%
}{\left[ 1-\left( 1-e^{-\alpha x_{i}-\gamma x_{i}^{\beta }}\right) ^{c}%
\right] },
\end{align}%
\begin{align}\label{eq6.7}
\frac{\partial \ell}{\partial \gamma } &=\sum\limits_{i=1}^{n}\frac{%
\beta x_{i}^{\beta -1}}{\left( \alpha +\gamma \beta x_{i}^{\beta -1}\right) }%
-\sum\limits_{i=1}^{n}x_{i}^{\beta }+(ac-1)\sum\limits_{i=1}^{n}\frac{%
x_{i}^{\beta }e^{-\alpha x_{i}-\gamma x_{i}^{\beta }}}{\left[ 1-e^{-\alpha
x_{i}-\gamma x_{i}^{\beta }}\right] }  \notag \\
&+(b-1)\sum\limits_{i=1}^{n}\frac{x_{i}^{\beta }e^{-\alpha x_{i}-\gamma
x_{i}^{\beta }}\left( 1-e^{-\alpha x_{i}-\gamma x_{i}^{\beta }}\right) ^{c-1}%
}{\left[ 1-\left( 1-e^{-\alpha x_{i}-\gamma x_{i}^{\beta }}\right) ^{c}%
\right] },
\end{align}%
and%
\begin{align}\label{eq6.8}
\frac{\partial \ell}{\partial \beta } &=\sum\limits_{i=1}^{n}\frac{%
\gamma x_{i}^{\beta -1}\left( \beta \log x_{i}+1\right) }{\left( \alpha
+\gamma \beta x_{i}^{\beta -1}\right) }-\gamma
\sum\limits_{i=1}^{n}x_{i}^{\beta }\log x_{i}  \notag \\
&+\gamma (ac-1)\sum\limits_{i=1}^{n}\frac{x_{i}^{\beta }\log
x_{i}e^{-\alpha x_{i}-\gamma x_{i}^{\beta }}}{\left[ 1-e^{-\alpha
x_{i}-\gamma x_{i}^{\beta }}\right] }  \notag \\
&+\gamma c(b-1)\sum\limits_{i=1}^{n}\frac{x_{i}^{\beta }\log
x_{i}e^{-\alpha x_{i}-\gamma x_{i}^{\beta }}\left( 1-e^{-\alpha x_{i}-\gamma
x_{i}^{\beta }}\right) ^{c-1}}{\left[ 1-\left( 1-e^{-\alpha x_{i}-\gamma
x_{i}^{\beta }}\right) ^{c}\right] }.
\end{align}%
By solving this nonlinear system of equations \eqref{eq6.3} - \eqref{eq6.8}, these solutions
will yield the ML estimators for $\widehat{a}$,$\widehat{b}$ , $\widehat{c}$
$,\widehat{\alpha }$, $\widehat{\beta }$ and $\widehat{\gamma }$ , for the
six parameters McDonald modified Weibull distribution $McMW$ $(\alpha
,\gamma ,\beta ,a,b,c,x)$ pdf all the second order derivatives exist. Thus
we have the inverse dispersion matrix is given by

\begin{equation}\label{eq6.9}
\left(
\begin{array}{c}
\widehat{\alpha } \\
\widehat{\theta } \\
\widehat{\beta } \\
\widehat{a} \\
\widehat{b} \\
\widehat{c}%
\end{array}%
\right) \sim N\left[ \left(
\begin{array}{c}
\alpha \\
\theta \\
\beta \\
a \\
b \\
c%
\end{array}%
\right) ,\left(
\begin{array}{cccccc}
\widehat{V_{\alpha \alpha }} & \widehat{V_{\alpha \theta }} & \widehat{%
V_{\alpha \beta }} & \widehat{V_{\alpha a}} & \widehat{V_{\alpha b}} &
\widehat{V_{\alpha c}} \\
\widehat{V_{\theta \alpha }} & \widehat{V_{\theta \theta }} & \widehat{%
V_{\theta \beta }} & \widehat{V_{\theta a}} & \widehat{V_{\theta b}} &
\widehat{V_{\theta c}} \\
\widehat{V_{\beta \alpha }} & \widehat{V_{\beta \theta }} & \widehat{%
V_{\beta \beta }} & \widehat{V_{\beta a}} & \widehat{V_{\beta b}} & \widehat{%
V_{\beta c}} \\
\widehat{V_{a\alpha }} & \widehat{V_{a\theta }} & \widehat{V_{a\beta }} &
\widehat{V_{aa}} & \widehat{V_{ab}} & \widehat{V_{ac}} \\
\widehat{V_{b\alpha }} & \widehat{V_{b\theta }} & \widehat{V_{b\beta }} &
\widehat{V_{ba}} & \widehat{V_{bb}} & \widehat{V_{bc}} \\
\widehat{V_{c\alpha }} & \widehat{V_{c\theta }} & \widehat{V_{c\beta }} &
\widehat{V_{ca}} & \widehat{V_{cb}} & \widehat{V_{cc}}%
\end{array}%
\right) \right] .
\end{equation}%
\begin{equation}\label{eq6.10}
V^{-1}=-E\left[
\begin{array}{cccccc}
V_{\alpha \alpha } & V_{\alpha \theta } & V_{\alpha \beta } & V_{\alpha a} &
V_{\alpha b} & V_{\alpha c} \\
&  &  &  &  &  \\
&  &  &  &  &  \\
&  &  &  &  &  \\
&  &  &  &  &  \\
V_{c\alpha } & V_{c\theta } & V_{c\beta } & V_{ca} & V_{cb} & V_{cc}%
\end{array}%
\right]
\end{equation}%
where
\begin{align*}
V_{\alpha \alpha } &=\frac{\partial ^{2}L}{\partial \alpha ^{2}},V_{\gamma
\gamma }=\frac{\partial ^{2}L}{\partial \gamma ^{2}},V_{\beta \beta }=\frac{%
\partial ^{2}L}{\partial \beta ^{2}} \\
V_{aa} &=\frac{\partial ^{2}L}{\partial a^{2}},V_{bb}=\frac{\partial ^{2}L}{%
\partial b^{2}},V_{cc}=\frac{\partial ^{2}L}{\partial c^{2}} \\
V_{a\alpha } &=&\frac{\partial ^{2}L}{\partial \alpha \partial a}%
,V_{_{\theta \beta }}=\frac{\partial ^{2}L}{\partial \theta \partial \beta }%
,V_{\beta \gamma }=\frac{\partial ^{2}L}{\partial \beta \partial \gamma }
\end{align*}%
By solving this inverse dispersion matrix these solutions will yield
asymptotic variance and covariances of these ML estimators for $\widehat{a}$,%
$\widehat{b}$ , $\widehat{c}$ $,\widehat{\alpha }$, $\widehat{\beta }$ and $%
\widehat{\gamma }$ . Using \eqref{eq6.9}, we approximate $100(1-\gamma )\%$
confidence intervals for $a,b,c,\alpha ,\beta ,$and $\gamma ,$are determined
respectively as
\begin{align*}
&\widehat{\alpha }\pm z_{\frac{\gamma }{2}}\sqrt{\widehat{V_{\alpha \alpha }%
}},\text{ }\widehat{a}\pm z_{\frac{\gamma }{2}}\sqrt{\widehat{V_{aa}}}\text{%
\ , }\widehat{\beta }\pm z_{\frac{\gamma }{2}}\sqrt{\widehat{V_{\beta \beta }%
}} \\
&\widehat{\gamma }\pm z_{\frac{\gamma }{2}}\sqrt{V_{\gamma \gamma }},%
\widehat{b}\pm z_{\frac{\gamma }{2}}\sqrt{\widehat{V_{bb}}}\text{ and }%
\widehat{c}\pm z_{\frac{\gamma }{2}}\sqrt{\widehat{Vcc}}
\end{align*}%
where $z_{\gamma }$ is the upper $100\gamma _{the}$ percentile of the
standard normal distribution.

\section{Application}

In this section, we use a real data set to show that the McMW distribution can be a better model than one based on
the MW  distribution and Weibull distribution. The data set given in Table 1 taken from \cite{libri} page 180 represents the failure times of 50 components(per 1000h):

\begin{table}[hbt]
 \caption{\label{Tab1} Failure Times of 50 Components(per 1000 hours).}
 \begin{small}
 \begin{center}
 \begin{tabular}{llllllllll}
 \hline
0.036&0.058&0.061&0.074&0.078&0.086&0.102&0.103&0.114&0.116\\
0.148&0.183&0.192&0.254&0.262&0.379&0.381&0.538&0.570&0.574\\
0.590&0.618&0.645&0.961&1.228&1.600&2.006&2.054&2.804&3.058\\
3.076&3.147&3.625&3.704&3.931&4.073&4.393&4.534&4.893&6.274\\
6.816&7.896&7.904&8.022&9.337&10.940&11.020&13.880&14.730&15.080\\
 \hline
 \end{tabular}
 \end{center}
 \end{small}
\end{table}

\begin{table}[hbt]
 \caption{\label{Tab1} Estimated parameters of the   Modified Weibull   and McMW  distribution for the failure times of 50 components(per 1000h).}
 \begin{small}
 \begin{center}
 \begin{tabular}{llll}
 \hline
 Model      & Parameter Estimate      & Standard Error & $-\ell(\cdot; x)$ \\
 \hline
 %Weibull & $\hat{\alpha} = 0.661$    & $0.074$       &  102.364\\
  %        &$\hat{\beta} = 2.531$     &0.571                    &      \\\hline
Modified  & $\hat{\alpha} = 0.043$    & $0.131$       & 102.320\\
   Weibull & $\hat{\beta} =  0.492$ &   0.181            &\\
            &  $\hat{\gamma}= 0.619 $&  0.154                &\\\hline

Mc Donald & $\hat{\alpha}= 0.599$ & 1.116e-05       &  98.404\\
modified & $\hat{\beta} = 1.063$ &  0.012      & \\
Weibull  &$\hat{\gamma}= 1.209$&  0.003          &          \\
      &$\hat{a}= 0.091 $ & 0.015         &          \\
    &$\hat{b}=  0.090 $ &  0.015           &          \\
    &$\hat{c}= 9.169 $ &   0.018         &          \\				
 \hline
 \end{tabular}
 \end{center}
 \end{small}
\end{table}

The variance covariance matrix $I(\hat\varphi)^{-1}$ of the MLEs under the McMW  distribution is computed as
$$
 \begin{pmatrix}
1.2\cdot 10^{-4}&-1.3\cdot 10^{-4}&6.2\cdot 10^{-6}&7.0\cdot 10^{-6}&1.2\cdot 10^{-5}&8.6\cdot 10^{-4}\\
-1.3\cdot 10^{-4}&1.4\cdot 10^{-4}&-1.1\cdot 10^{-5}&-7.6\cdot 10^{-6}&-1.2\cdot 10^{-5}&-9.3\cdot 10^{-4}\\
6.2\cdot 10^{-6}&-1.1\cdot 10^{-5}&9.5\cdot 10^{-6}&6.0\cdot 10^{-7}&7.2\cdot 10^{-7}&6.4\cdot 10^{-5}\\
7.0\cdot 10^{-6}&-7.6\cdot 10^{-6}&6.0\cdot 10^{-7}&2.4\cdot 10^{-4}&8.6\cdot 10^{-5}&6.5\cdot 10^{-5}\\
1.2\cdot 10^{-5}&-1.2\cdot 10^{-5}&7.2\cdot 10^{-7}&8.6\cdot 10^{-5}&2.4\cdot 10^{-4}&8.4\cdot 10^{-5}\\
8.6\cdot 10^{-4}&-9.3\cdot 10^{-4}&6.4\cdot 10^{-5}&6.5\cdot 10^{-5}&8.4\cdot 10^{-5}&3.5\cdot 10^{-3}\\
\end{pmatrix}\,.
$$
Thus, the variances of the MLE of $\alpha, \beta, \gamma, a, b $ and $c$ is 
$var(\hat\alpha) = 1.247\cdot 10^{-4},var(\hat \beta) = 1.488\cdot 10^{-4},$ $var(\hat \gamma) =  9.529\cdot 10^{-6},
var(\hat  a)=2.472\cdot 10^{-4},$ $var(\hat b) =2.467\cdot 10^{-4},var(\hat c) =3.544\cdot 10^{-3}.$

Therefore, $95\%$ confidence intervals for $\alpha, \beta, \gamma, a, b$ and $c$ are $[0.599,0.601],[1.039,1.086], [ 1.202, 1.215],[0.06, 0.121],[ 0.059,0.121] $ and $[9.132,  9.205]$ respectively.

\begin{table}[hbt]
 \caption{\label{Tab2} Criteria for comparison.}
 \begin{center}
 \begin{small}
 \begin{tabular}{llllllll}
 \hline
  Model           K-S  & $-2\ell$ & AIC     & AICC    \\
  \hline
MW             0.128&   204.640 &210.64  &211.161  \\
McMW             0.118& 196.808 & 208.808  &210.761  \\
  \hline
 \end{tabular}
 \end{small}
 \end{center}
\end{table}

In order to compare the two distribution models, we consider criteria like , $-2\ell$, AIC (Akaike information criterion)and AICC (corrected Akaike information criterion)  for the data set. The better distribution corresponds to smaller  $-2\ell$, AIC and AICC  values:`
$$
 \mbox{AIC} = 2k - 2\ell\,, \quad \textrm{and}
 \mbox{AICC} = \mbox{AIC} + \frac{2k(k+1)}{n-k-1}\,,
$$

where $k$ is the number of parameters in the statistical model, $n$  the sample size and $\ell$ is the maximized value of the log-likelihood function under the considered model. Table 3 shows the MLEs under both distributions, table 4 shows the values of  K-S, $-2\ell$, AIC and AICC values. The values in table 3 indicate that the McMW distribution leads to a better fit than the  MW  distribution and Weibull distribution.

A density plot compares the fitted
densities of the models with the empirical histogram of the observed data (Fig. 4).
The fitted density for the McMW model is closer to the empirical histogram than the
fits of the MW and Weibull sub-models.
\begin{figure}[H]
\centering
 \vspace*{-20mm}
  \includegraphics[width=8.0cm]{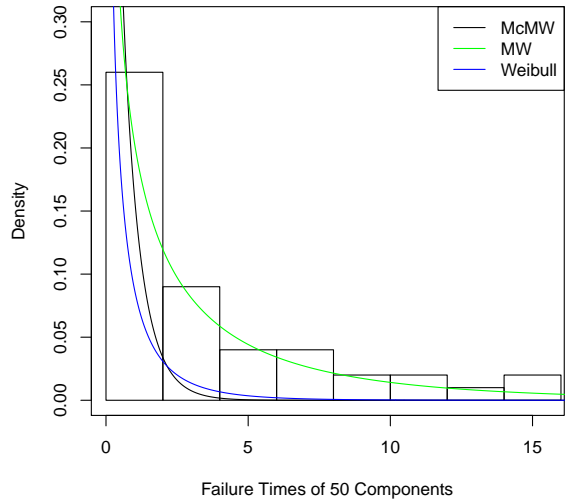}
 \caption{Estimated densities of the models for  the failure times of 50 components.\label{fig4.pdf}}
 \end{figure}

\begin{figure}[H]
 \vspace*{-20mm}
 \begin{center}
\includegraphics[width=8cm]{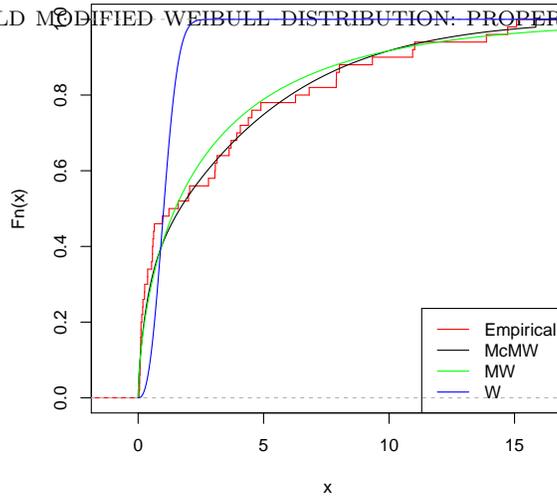}
\caption{Empirical, fitted Weibull, MW and McMW cdf of the failure times of 50 components. \label{figaaaaaq.pdf}}
\end{center}
\end{figure}

\section{Conclusion}

Here we propose a new model, the so-called the Mc Donald Modified Weibull distribution  which extends the
modified Weibull  distribution in the analysis of data with real support. An obvious reason for generalizing a standard distribution is because the generalized form provides larger flexibility in modeling real data. We derive expansions for  moments and for the moment generating function. The estimation of parameters is approached by the method of maximum likelihood, also the information matrix is derived.  An application of McMW distribution to real data show that the new distribution can be used quite effectively to provide better fits than modified Weibull and  Weibull distribution.


\begin{thebibliography}{99}
\bibitem{1} Aarset, M. V. (1987). How to identify a bathtub hazard rate. Reliability, IEEE Transactions on, 36(1), 106-108.
\bibitem{2} Bain, L. J. (1974). Analysis for the linear failure-rate life-testing distribution. Technometrics, 16(4), 551-559.
\bibitem{3} Barlow R. E.,  and Proschan F. (1981). Statistical Theory of Reliability and Life Testing, Begin
With, Silver Spring, MD,.
\bibitem{carrasco} Carrasco, J. M., Ortega, E. M., and Cordeiro, G. M. (2008). A generalized modified Weibull distribution for lifetime modeling. Computational Statistics and Data Analysis, 53(2), 450-462.
\bibitem{4} Eddy, O. N. (2007). Applied statistics in designing special organic mixtures. Applied Sciences, 9, 78-85.
\bibitem{grad} Gradshteyn, I. S., and Ryzhik, I. M. (2000). Table of Integrals, Series, and Products 6th edn (New York: Academic).
\bibitem{5} Ghitany, M. E., and Kotz, S. (2007). Reliability properties of extended linear failure-rate distributions. Probability in the Engineering and informational Sciences, 21(3), 441.
\bibitem{6} Lawless J. F. (2003). Statistical Models and Methods for Lifetime Data, John Wiley and Sons,
New York.
\bibitem{lee} Lee, C., Famoye, F., and Olumolade, O. (2007). Beta-Weibull distribution: some properties and applications to censored data. Journal of modern applied statistical methods, 6(1), 17.
\bibitem{7} Lin, C. T., Wu, S. J., and Balakrishnan, N. (2006). Monte Carlo methods for Bayesian inference on the linear hazard rate distribution. Communications in Statistics—Simulation and Computation, 35(3), 575-590.
\bibitem{8} Miller Jr, R. G. (2011). Survival analysis (Vol. 66). John Wiley and Sons.
\bibitem{libri} Murthy, D. P., Xie, M., and Jiang, R. (2004). Weibull models (Vol. 505). John Wiley and Sons.
\bibitem{silva} Silva, G. O., Ortega, E. M., and Cordeiro, G. M. (2010). The beta modified Weibull distribution. Lifetime data analysis, 16(3), 409-430.
\bibitem{9} Tadj, L., Sarhan, A. M., and El-Gohary, A. (2008). Optimal control of an inventory system with ameliorating and deteriorating items. Applied Sciences, 10, 243-255.
\bibitem{10} Zhang, Z., Sun, H., and Zhong, F. (2007). Information geometry of the power inverse Gaussian distribution. Applied Sciences (APPS), 9, 194-203.
\end{thebibliography}
\end{document}